# Computational Comparison of Optimal Experimental Design Strategies for Two- and Three-Dimensional Lung Electrical Impedance Tomography

**Corneliu Arsene[1,2]**

[1]*School of Mathematics, University of Manchester, UK*

[2]*Centre for Advanced Research and Innovation, Romania*

*Email: corneliuarsene@gmx.com , arsenecorneliu@tutanota.de*

**Abstract**

Optimal experimental design (OED) in electrical impedance tomography (EIT) seeks informative current-injection and voltage-measurement configurations for conductivity reconstruction. This paper investigates two computational strategies for two- and three-dimensional lung EIT. Algorithm 1 minimises an empirical Bayesian reconstruction risk through bilevel optimisation. In a 2D numerical example involving 200 iterations and 45 samples, parallel MATLAB execution on three Broadwell HPC nodes equipped with Intel Xeon E5-2680 v4 processors reduced the computational time from 944 to 39 minutes, achieving an observed 24.2-fold speedup. Algorithm 2 addresses 3D EIT through a two-stage optimisation procedure. First, optimal current-drive patterns are determined by evaluating their ability to distinguish simulated inhalation and exhalation conductivity states, leading to the selection of 16 informative electrodes from an initial configuration of 64 electrodes arranged in four rings. Second, linearly independent voltage measurements are selected using a procedure adapted from [7]. The resulting measurement protocols are further evaluated using the truncated condition number of the transfer-impedance Jacobian. Numerical reconstructions of a cylindrical model containing two lung-like inclusions demonstrate that the optimised measurement protocol preserves reconstruction quality comparable to that obtained with a conventional protocol. The study highlights the potential of computational optimisation for identifying informative measurement configurations in 2D and 3D lung EIT.

## 1. Introduction

Electrical impedance tomography (EIT) reconstructs conductivity from boundary currents and voltages. Lung EIT provides non-ionising, rapid monitoring, but reconstruction remains ill-posed and sensitive to electrode geometry and measurement choice [1–4,9,36]. Optimal experimental design (OED) therefore seeks current and voltage protocols that convey useful information without excessive computational or acquisition cost.

Relevant foundations include Bayesian-risk design for ill-posed inverse problems [5,6], distinguishability-based current patterns and their safety constraints [34,35], independent-measurement selection [7], and electrode-position optimisation [8]. Later work has explored alternative Bayesian objectives, data-driven placement and 3D acquisition arrangements [11,12,15–18,23]. These approaches optimise different criteria and are not directly interchangeable.

Related papers/contributions investigated transient-aware voltage estimation with discrete and 3D phantoms [37,38] and described a 64-channel system for exploring EIT acquisition protocols [39]. These developments provide the experimental context for the present computational study, which uses simulated voltage data rather than physical measurements.

This paper compares two computational approaches: parallel bilevel Bayesian-risk minimisation for 2D EIT, and a two-stage strategy for 3D lung EIT that ranks current-drive electrodes from simulated inhalation–exhalation impedance contrasts before selecting linearly independent voltage measurements. The second stage adapts the independence procedure of [7], whereas the combined contrast-weighted electrode ranking and measurement-selection sequence is the particular implementation examined here.

The discussion focuses on numerical reconstruction, computational cost and Jacobian conditioning. Saline-phantom real experiments are not included in this computational study.

## 2. Computational methods

### 2.1 Forward model and Bayesian-risk optimisation

For conductivity $\sigma$ in a domain $\Omega$, the quasistatic forward problem satisfies $\nabla\cdot(\sigma\nabla u) = 0$ with electrode boundary conditions. In discrete form, transfer-impedance data are represented as $y = F(\sigma; d) + \varepsilon$, where d denotes the acquisition protocol, $F(\sigma; d)$ is the EIT forward model that predicts electrode voltages for a given conductivity distribution and experimental design and $\varepsilon$ is measurement error. The Jacobian $J(d) = \partial F/\partial\sigma$ measures local sensitivity to conductivity changes [1,2]. Related estimation problems occur in hydraulic distribution networks, where loop-flow modelling, least-squares state estimation, confidence-limit analysis and decision support have been investigated [24,30–33]. These studies concern hydraulic network states rather than EIT conductivity; the broader mathematical connection is estimation under incomplete or uncertain observations [25,26].

Algorithm 1 treats 2D protocol selection as bilevel optimisation: the outer problem seeks a design d that reduces the expected loss of an inner reconstruction. In the empirical Bayesian-risk formulation [5,6],

$$d^* \in \arg\min R(d), \qquad R(d) = E_{\sigma,\varepsilon}\ \{\ L[\sigma, \hat{\sigma}(F(\sigma; d) + \varepsilon; d)]\ \}. \qquad (1)$$

where d is experimental design, specifying the current-injection and voltage-measurement configurations, $d^*$ is optimal experimental design that minimises the Bayesian risk, $R(d)$ is Bayesian risk associated with experimental design d, $\sigma$ is true electrical conductivity distribution inside the investigated domain, $\boldsymbol{\varepsilon}$ is measurement noise affecting the electrode voltages, $F(\sigma; d)$ is EIT forward model that predicts electrode voltages for a given conductivity distribution and experimental design, $F(\sigma; d) + \varepsilon$ is simulated electrode voltages including measurement noise, $\hat{\sigma}$ is reconstructed electrical conductivity distribution obtained from the measurements, $L(\sigma, \hat{\sigma})$ is loss function measuring the difference between the true and reconstructed conductivity distributions, $E_{\sigma,\varepsilon}$ is expected value over possible conductivity distributions and measurement-noise realizations, arg min is set of experimental designs that minimise the Bayesian risk.

Algorithm 1 entails repeated forward and inverse solutions across conductivity and noise samples. The 2D example used 200 outer iterations and 45 samples. MATLAB *parfor* was distributed over three Broadwell nodes (independent HPC servers with Intel Xeon E5-2680 v4 processors, 2.40 GHz, and 128 GB RAM each) at Manchester's Computational Shared Facility. One node coordinated parameter updates using four cores, while two nodes processed sample batches using eight cores each. Related computational developments include EIT numerical software and high-performance probabilistic modelling [27–29].

The reconstruction $\hat{\sigma}$ and loss L in the risk objective are evaluated repeatedly for each sampled candidate design, making the computational cost substantial.

## 2.2 Algorithm 2: contrast-weighted selection of electrodes and independent 3D measurements

Algorithm 2 determines a 3D current-injection and voltage-measurement protocol in two stages. Optimal-current theory provides candidate drives [34,35]; electrodes are then ranked using the simulated inhalation–exhalation response before independent voltage observations are selected by a procedure based on [7]. Here $N_e$ denotes available electrodes and $M_e$ the number selected.

Algorithm 2 — Two-stage optimal protocol selection for 3D EIT

1. Generate candidate current-drive patterns using a distinguishability criterion for the simulated lung conductivity states, subject to admissible current constraints [34,35].
2. Evaluate candidate electrodes from the simulated difference in transfer impedance between inhalation and exhalation; rank their contributions and retain the $M_e$ most informative electrodes out of $N_e$.
3. For the selected drives, determine the optimal set of linearly independent voltage measurements using the independence-based selection procedure adapted from [7].
4. Assemble the measurement protocol and assess its numerical conditioning using the truncated singular-value decomposition of the transfer-impedance Jacobian.

For a selected Jacobian J, the optional truncated condition number $\kappa_{ns}$, based on its first $n_s$ singular values $s_1 \geq \ldots \geq s_{ns} > 0$, is

$$\kappa_{ns}(J) = s_1 / s_{ns}. \quad (2)$$

A smaller value indicates a better-conditioned measurement protocol, whereas a larger value indicates greater sensitivity to measurement errors and numerical instability. Therefore, this condition number evaluates protocol conditioning. The numerical examples also examined current-pair separation and the spacing and orientation of electrode patches.

The methodological contribution considered here is the integration of lung-state-based electrode ranking and independent voltage-measurement selection, rather than the invention of optimal-current theory or linear-independence testing [7,34,35].

# 3. Computational results

## 3.1 Two-dimensional optimisation

For 200 iterations and 45 samples, the desktop required 944 minutes (approximately 16 hours) and the three Broadwell nodes (Xeon E5-2680 v4 compute servers, 128 GB RAM each) required 39 minutes, an observed 24.2-fold wall-clock speedup. Runs with eight desktop MATLAB workers and one Broadwell node took 359 and 62.81 minutes, respectively.

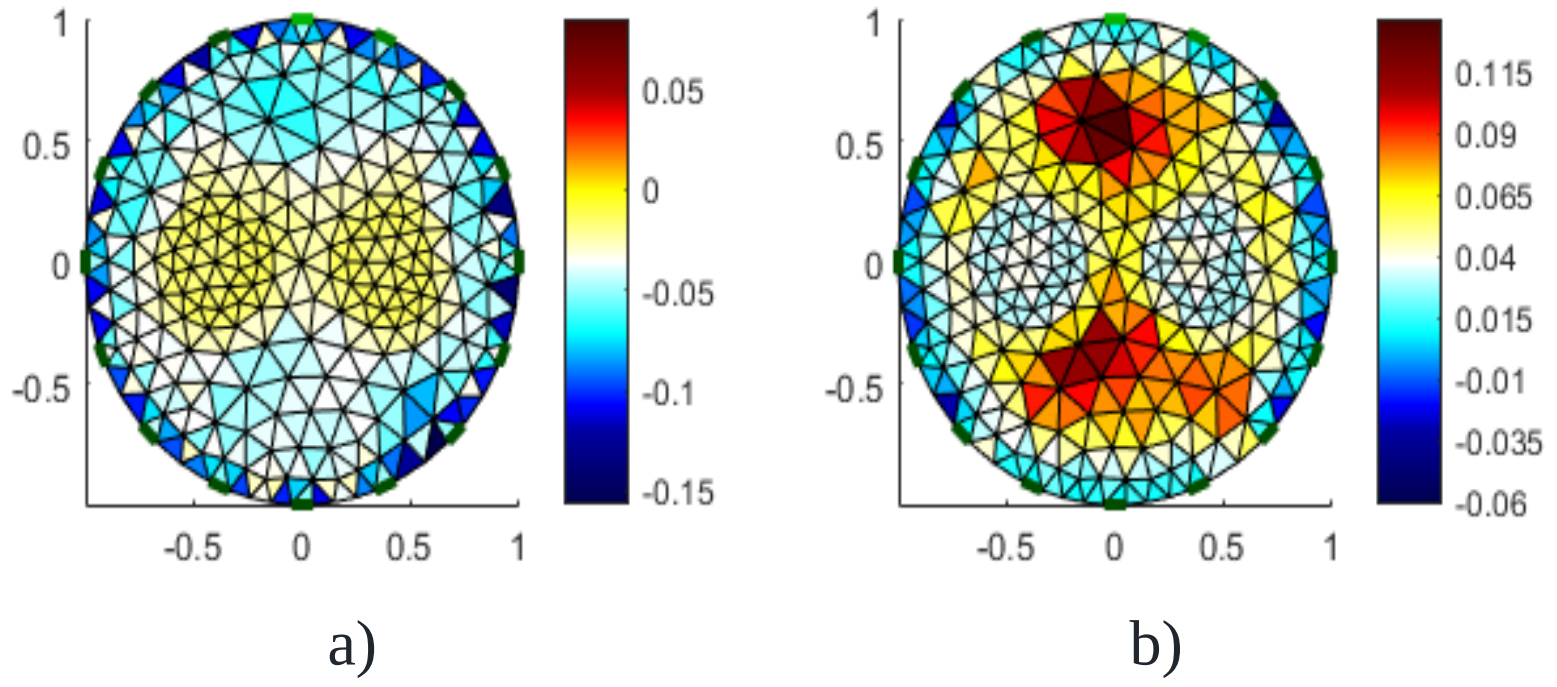

**Figure 1:** Comparison between the conductivity images calculated from the sensitivities of the most important 46 measurements (a) versus the conductivity images obtained from sensitivities of least important 90 measurements (b).

### 3.2 Three-dimensional protocol selection

For the cylindrical 3D model, the contrast-weighted ranking selected $M_e = 16$ electrodes from $N_e = 64$. Candidate current drives were assessed for their ability to distinguish simulated inhalation and exhalation states, and a set of linearly independent voltage observations was subsequently selected. Numerical conditioning and reconstruction quality were then evaluated for the resulting protocol.

The numerical reconstruction using the optimised measurement pattern is presented in Figure 2. The original 3D model contains two lung-like inclusions and 64 electrodes arranged in four rings of 16, from which Algorithm 2 identified 16 informative electrodes. The reconstructions use simulated voltage data rather than saline-phantom measurements. The optimised measurement protocol successfully reconstructed the simulated conductivity distribution, demonstrating that the proposed selection procedure can identify informative current-drive and voltage-measurement configurations while maintaining reconstruction quality comparable to that of a conventional protocol.

The 3D simulations also compared conductivity and voltage discrepancies through $L_2$ norms over several regularisation hyperparameters. Their purpose was to assess whether the selected measurements preserved useful reconstruction information, not to demonstrate universal superiority over other EIT protocols.

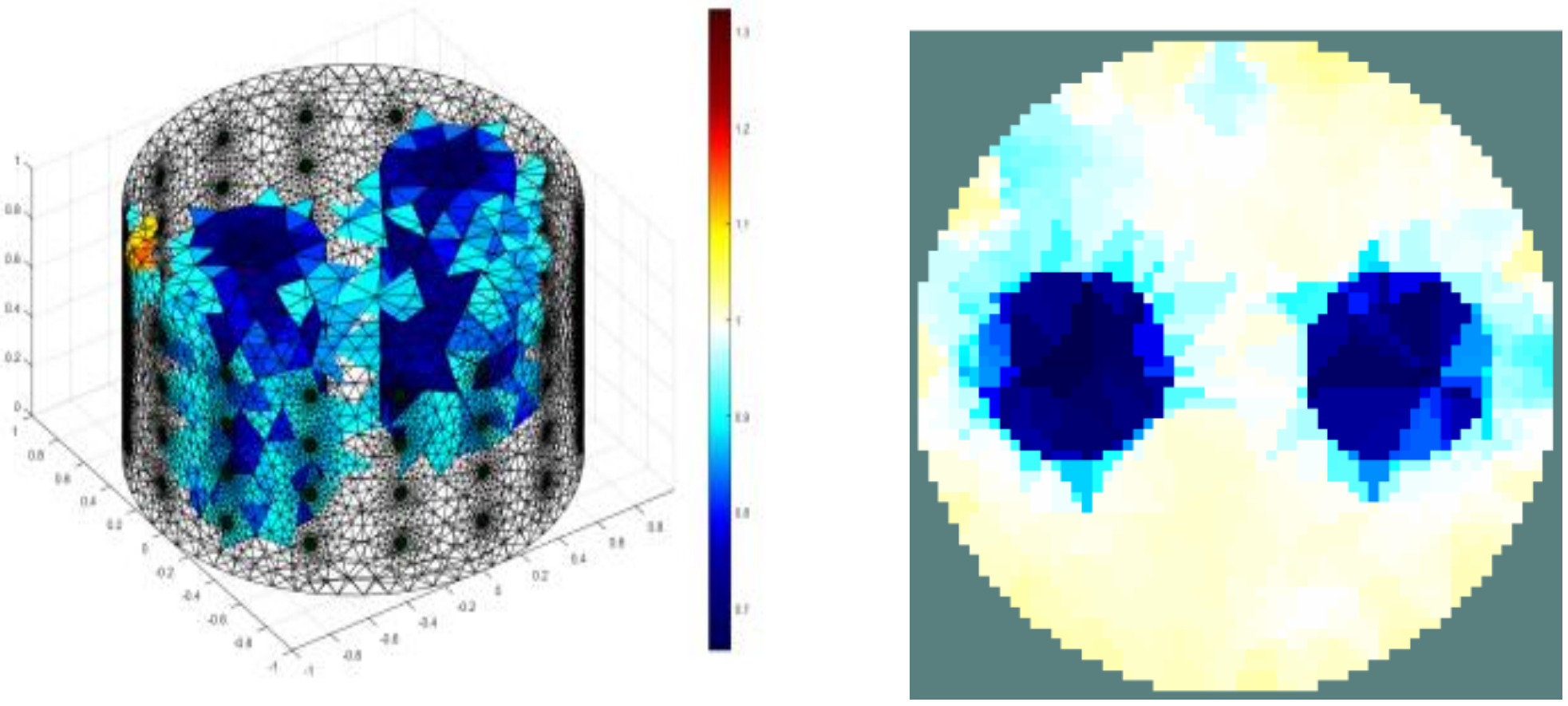

a) b)

**Figure 2.** Numerical conductivity reconstruction using the optimal measurement protocol determined by Algorithm 2, which selected 16 informative electrodes from an initial configuration of 64 electrodes arranged in four rings of 16 around a cylindrical 3D EIT model. The reconstruction uses simulated voltage measurements: (a) reconstructed cylinder; (b) transverse slice.

### 4. Relation to established and contemporary EIT methods

Bayesian and risk-based criteria. Haber et al. [5,6] provided the optimisation framework for ill-posed inverse problems underlying Algorithm 1. Karimi et al. [12] subsequently used expected information gain, while Hyvönen et al. [16] applied Bayesian design to head EIT. These criteria differ from empirical reconstruction-risk minimisation. The present 2D result concerns the computational cost of repeated risk evaluations.

Electrode placement and model uncertainty. The importance of electrode geometry was recognised in earlier 3D probe design [7] and in the optimisation of electrode positions [8]. Later investigations considered deep-learning-based placement [11], domain parameters [13] and model robustness [10,14,22]. Against this background, the four-ring cylindrical model provides a controlled numerical example.

Excitation and measurement selection. The optimal-current-pattern theory of Gisser et al. [35] and the safety-constrained formulation of Lionheart et al. [34] underpin the first-stage drive candidates. Algorithm 2 then introduces a lung-state contrast score for ranking individual electrodes before independent voltage observations are chosen following Borsic et al. [7]. The combination differs in objective and sequence from studies of scanning electrodes [15], staged 3D configurations [23] and alternative excitation schemes [17,18].

Relationship to the R3M studies. Earlier studies of transient-aware voltage estimation [37,38] and 64-channel EIT instrumentation [39] provide complementary experimental motivation. The present paper focuses on computational measurement-protocol selection using simulated data.

Assessment criteria. The general mathematical challenges of impedance imaging are reviewed in [36]. In the present study, Jacobian singular-value decay and the truncated condition number are employed to evaluate and compare the numerical conditioning of different measurement protocols. Together with the conductivity reconstruction results, these analyses provide complementary evidence supporting the effectiveness of the proposed measurement-selection strategy for 3D EIT. GREIT [3], clinical guidance [4,19,20] and sensitivity-enhancement studies [21] address other criteria. Fair comparison requires common measurement budgets, noise models and reconstruction methods.

For risk-based methods, computational cost, priors and reconstruction loss must also be controlled. The present 2D and 3D simulations examine complementary approaches rather than a matched head-to-head benchmark.

## 5. Discussion

Algorithm 1 optimises empirical Bayesian risk through repeated inverse reconstructions; Algorithm 2 instead assigns electrode scores from differences between simulated lung states, selects preferred current drives and filters voltage measurements for independence. The first uses a distributed cluster to accelerate an expensive objective; the second uses a structured selection sequence for the 3D model.

The methodological contribution of Algorithm 2 lies in integrating lung-state-contrast-weighted electrode ranking with the selection of linearly independent voltage measurements to optimise measurement protocols for 3D lung EIT. While optimal current patterns [34,35], independent-measurement selection [7] and staged electrode-configuration optimisation [23] have been investigated previously, the present study examines their integration within a unified computational procedure for a 64-electrode lung EIT model. The numerical conditioning of the resulting measurement configurations is additionally assessed through truncated Jacobian singular-value analysis.

In the reported 3D example, 16 electrodes were retained and the resulting measurement pattern produced reconstructions visually comparable to those from a conventional protocol. The 2D benchmark establishes a substantial observed reduction in wall-clock time; the examples do not constitute a matched quantitative comparison between the two algorithms.

## 6. Conclusions

The paper has examined two complementary EIT design strategies. Parallel Bayesian-risk optimisation for 2D EIT reduced computation from 944 to 39 minutes on the reported configurations. For simulated 3D lung EIT, a two-stage procedure combined distinguishability-based current drives, lung-state-contrast-weighted ranking of 16 out of 64 electrodes, and selection of independent voltage measurements, with truncated Jacobian conditioning as a supplementary assessment. Its numerical reconstructions were comparable in visual quality to a conventional protocol. Real physical saline-phantom studies undertaken are outside the present paper.

## Acknowledgement

This work was supported by the UK Engineering and Physical Sciences Research Council (EPSRC) through the project Robust Repeatable Respiratory Monitoring with EIT (R3M), grant EP/L019108/1, and access to computational resources at the University of Manchester.